# Comments on the paper: 'Crystal growth and comparison of vibrational and thermal properties of semi-organic nonlinear optical materials'

Bikshandarkoil R Srinivasan, Royle Fernandes
Department of Chemistry, Goa University, Goa 403206, INDIA
Email: srini@unigoa.ac.in Telephone: 0091-(0)832-6519316; Fax: 0091-(0)832-2451184

**Abstract**

Gunasekaran et al (*Pramana – J. Phys.* **75** 683-690 (2010)) report to have grown the nonlinear optical crystals urea thiourea mercuric chloride (UTHC) and urea thiourea mercuric sulphate (UTHS). We argue that UTHC and UTHS are dubious crystals and are not what the authors propose.

**Comment**

During the course of a literature survey of metal compounds containing both thiourea and urea ligands, the title paper by Gunasekaran et al [1] reporting on growth of so called urea thiourea mercuric chloride (UTHC) and urea thiourea mercuric sulphate (UTHS) crystals attracted our attention. For formulating these crystals, the authors did not take into consideration the known chemistry of Hg(II) towards thiourea [2]; instead assumed that the mixing of urea, thiourea and mercuric chloride (or sulphate) in water will result in the formation of the so called UTHC (or UTHS) crystals by slow evaporation method. This assumption can be evidenced by the fact that i) both compounds are not represented by proper chemical formula but by unusual names not in accordance with chemical nomenclature and abbreviated by strange codes UTHC and UTHS and ii) no results of single crystal X-ray structure determination and chemical analysis are given in the entire paper to support the exact composition of these so called UTHC and UTHS crystals. Interestingly the authors claimed to have used atomic absorption spectroscopy (AAS) to determine the mole percentage of dopants incorporated in their grown doped crystals, which only adds confusion because in the synthetic details of crystal growth there is no mention of growth of any doped crystals other than the so called UTHC and UTHS. Further the exact nature of the dopant cannot also be inferred from the following claim of the author which can only be termed as unfortunate. *The low percentage of incorporation of dopants into the crystal may be because of the large difference between the ionic radii*. The UV-visible spectral data do not provide any clue on the exact nature of the so called

UTHC and UTHS since all the reagents used in the crystal growth namely urea, thiourea, mecrcuric chloride (or sulphate) are colourless solids and are known to be transparent in the visible region.

The authors claim to have used infrared (IR) spectroscopic studies to identify the functional groups present in the crystals and to determine the molecular structure. Urea represented by the formula $CO(NH_2)_2$ exhibits an intense band at 1675 cm$^{-1}$ for the carbonyl (CO) vibration, which is absent in the reported IR spectra of UTHC and UTHS and also in the bands assigned by authors. Thus the IR spectra serve to infer the absence of urea in both UTHC and UTHS proving their dubious nature. This is not surprising and can be explained as the expected behaviour of a softer Hg(II) ion preferring to bind to the soft S donor of thiourea rather than the hard O of urea. This explanation gains credence from the fact that no example of a well characterized Hg(II) compound containing both urea and thiourea ligand in the same compound is reported in the literature unlike a series of mercury-thiourea complexes described half a century ago [2]. An earlier claim of the growth of a so called thiourea urea zinc sulphate (TUZS) crystal has been shown to be erroneous and the TUZS was correctly formulated as the well known tris(thiourea)zinc sulphate $[Zn(tu)_3(SO_4)]$ (tu = thiourea) [3].

In order to verify if any Hg(II) compound containing both urea and thiourea can be isolated from an aqueous medium by reaction of urea, thiourea and a mercuric salt we reinvestigated the crystal growth reactions under identical conditions in the title paper and compared the IR spectra and powder pattern of the product (so called UTHC and UTHS) thus obtained with those of reported Hg(II)-thiourea complexes. The perfect matching of the spectra and powder pattern of the so called UTHC (or UTHS) with those of $[Hg(tu)_2Cl_2]$ (or $[Hg(tu)_2(SO_4)]$) reveal i) the crystals reported in the title paper are dubious and ii) confirm unambiguously that they do not contain any urea. The present comment thus points out the risk of formulating new compounds based on an incorrect assumption that mixing of a few reagents in aqueous solution will result in the crystallization of a desired product.

## References

[1] S Gunasekaran, G Anand, R Arun Balaji, J Dhanalakshmi and S Kumaresan, *Pramana – J. Phys.* **75** 683-690 (2010).

[2] I Aucken and R S Drago *Inorganic Syntheses*, **6** 26-30 (1960).
http://onlinelibrary.wiley.com/doi/10.1002/9780470132371.ch9/summary

[3] B R Srinivasan, T A Naik, Z Tylczyński and K R Priolkar, *Spectrochim Acta* **A117** 805-809 (2014).
http://dx.doi.org/10.1016/j.saa.2013.08.083